\documentclass[a4paper]{jpconf}

\usepackage{graphicx}

\newcommand{\beq}{\begin{equation}}
\newcommand{\eeq}{\end{equation}}
\def\p{{\vec p}}

\def\d3p{\frac{d^3\p}{(2\pi)^3}E_\p}

\begin{document}
\title{Thermalization of the quark-gluon plasma and dynamical formation of Bose-Einstein Condensate}

\author{Jinfeng Liao}

\address{Physics Department and Center for Exploration of Energy and Matter,
Indiana University, 2401 N Milo B. Sampson Lane, Bloomington, IN 47408, USA.}

\address{RIKEN BNL Research Center, Bldg. 510A, Brookhaven National Laboratory, Upton, NY 11973, USA.}

\ead{liaoji@indiana.edu}

\begin{abstract}
We report recent progress on understanding the thermalization of the quark-gluon plasma during the early stage in a heavy ion collision. 
The initially high overpopulation in the pre-equilibrium gluonic matter (``glasma'') is shown to play a crucial role. The strongly interacting nature (and thus fast evolution) naturally arises as an {\em emergent property} of this pre-equilibrium matter where the  intrinsic coupling is weak but the highly occupied
gluon states coherently amplify the scattering. A possible transient Bose-Einstein Condensate is argued to form dynamically on a rather general ground. We develop the kinetic approach for describing this highly overpopulated system and find approximate scaling solutions as well as numerically study the onset of condensation. Finally we discuss possible phenomenological implications. 
\end{abstract}

\section{Introduction}

Thermalization of the quark-gluon plasma is one of the least understood and most challenging problems in current heavy ion physics.  The conception of the pre-equilibrium gluonic matter (``glasma'') is strongly constrained by both ends in the time evolution. Prior to forming this gluonic matter, the colliding nuclei are in a form of color glass condensate with high gluon occupation $1/\alpha_{\rm s}$ below saturation scale $Q_{\rm s}$ and following the collision there is a subsequent strong field evolution stage (likely also instabilities) till about the time $1/Q_{\rm s}$ and then succeeded by the glasma stage.   
The end of this glasma evolution toward a locally equilibrated quark-gluon plasma (QGP) is also strongly indicated by phenomenology to be reached on the order of a fermi over c time. With these constraints in mind, we explore here a thermalization scenario of strongly interacting matter emergent from weak coupling but large aggregate of constituents based on recent works in \cite{Blaizot:2011xf,Chiu:2012ij,BLM}. Let us also mention that this topic is currently under intensive investigations from a variety of approaches, see e.g. \cite{Kurkela:2011ti,Kurkela:2011ub,Epelbaum:2011pc,Gelis:2011xw,Berges:2012us,Berges:2011sb,Berges:2012ev,Berges:2012mc,Kurkela:2012hp,Schlichting:2012es,Berges:2012ks}.

\section{Highly overpopulated glasma}

With the thermalization problem in mind, let us start by considering the somewhat idealized problem of the evolution in a weakly coupled gluon system that is initially far from equilibrium and described by the following (glasma-type) distribution (with coupling $\alpha_{\rm s}<<1$): 
\begin{eqnarray} \label{eq_f_glasma}
f(p \leq Q_{\rm s}) = 1 / \alpha_{\rm s} \quad , \quad f(p>Q_{\rm s}) = 0
\end{eqnarray} 
The most important feature of this initial gluon system, as identified and emphasized in \cite{Blaizot:2011xf}, is the high overpopulation $1/\alpha_{\rm s}$. Three important consequences follow from this feature by very general arguments and may hold the key of understanding the thermalization process. 

\subsection{The emergence of a strongly interacting matter}

Different from the usual particle scattering cross section in the few-particle case, the scattering rate in the many-body setting receives factors from the occupation number in phase space both for initial and final states involved in the scattering. When the occupation number reaches a value as high as $1/\alpha_{\rm s}$, the power counting in coupling constant is necessarily rendered different from normal perturbation theory. For example let us consider the contribution from the $2\leftrightarrow 2$ gluon scattering process to the collision integral for a transport equation of gluon distribution $f(\vec p)$. Normally at weak coupling, this process contributes at order $\sim \hat{o} (\alpha_{\rm s}^2)$ and therefore is considered rather ``slow'' in bringing the system back to equilibrium. However in the ``glasma counting'' with $f\sim 1/ \alpha_{\rm s}$, there will be two factors from distribution functions and the resulting collision term shall be counted as $\sim \alpha_{\rm s}^2 f^2 \sim \hat{o}(1)$, despite how small the coupling $\alpha_{\rm s}$ may be! That is to say, the superficially strongly interacting nature in the pre-equilibrium stage (as implied by fast evolution toward equilibrium) is an {\em emergent property} of the weakly coupled albeit highly overpopulated glasma.   In passing we note that  this quite general argument can be extended to a large class of higher order scattering terms that all contribute at order $\hat{o}(1)$.

\subsection{From overpopulation to Bose-Einstein Condensation}

An even more drastic implication of the high overpopulation is that there are so many more gluons in the glasma as compared with a thermalized plasma for the same amount of energy that a Bose-Einstein Condensation(BEC) has to occur. To quantify this statement, let us introduce the overpopulation parameter as a dimensionless combination of the particle number density and energy density: $ n\, \epsilon^{-3/4}$. For the glasma distribution in Eq.(\ref{eq_f_glasma}) one has
\begin{eqnarray}
n \sim Q_{\rm s}^3 / \alpha_{\rm s} \quad , \quad \epsilon \sim Q_{\rm s}^4  \alpha_{\rm s} \quad , \quad n\, \epsilon^{-3/4} \sim 1/  \alpha_{\rm s}^{1/4} 
\end{eqnarray}
while for a thermal Bose gas one has 
\begin{eqnarray}
n \sim T^3  \quad , \quad \epsilon \sim T^4    \quad , \quad n\, \epsilon^{-3/4} \sim 1 
\end{eqnarray}
A more precise calculation shows that for a massless Bose gas $n\epsilon^{-3/4}=\frac{30^{3/4}\, \zeta(3)}{\pi^{7/2}} \approx 0.28$ (and for the massive case the number differs only slightly). Evidently as long as the coupling in the glasma distribution satisfies $\frac{1}{\alpha_{\rm s}} >  \frac{3^7\, 5^3\, \zeta(3)^4}{4\, \pi^{12}} \approx 0.154$ (which is easily achieved) then the initial $n\epsilon^{-3/4}$ is greater than the thermal value. For realistic $\alpha_{\rm s}$ values  $0.2\sim 0.3$   the system is significantly overpopulated. What is the consequence of that? Well, if the evolution of this system is dominated by elastic processes (at least over a certain time window), then both the particle number and energy will be conserved and therefore so is the combination $ n\, \epsilon^{-3/4}$. In such a case it is {\em impossible} for the system to simply equilibrate to a thermal Bose distribution $f_{eq}(\vec p)=1/(e^{\omega_{\vec p}/T}-1)$ with any values of mass and temperature. What will happen is that all the excessive gluons with the given amount of energy will have to be absorbed into a Bose-Einstein Condensate (BEC) that carries the no kinetic energy or momentum but as many particles as needed be. That is, the overpopulated system will thermalize to a distribution like $f_{eq}(\vec p)= n_c (2\pi)^3\delta^3(\vec p) + 1/(e^{\omega_{\vec p}/T}-1)$. In the glasma example one expects eventually a condensate with the density parametrically being $n_c \sim (Q_{\rm s}^3/\alpha_{\rm s}) (1-1/\alpha_{\rm s}^{1/4})$. This observation is quite fascinating: it took many years of efforts for atomic physicists to successfully reduce the energy in the confined cloud of Bose atoms (with roughly fixed total number in the cloud) enough thus bringing the system into overpopulation and BEC, while in heavy ion collisions the initial gluonic matter is born with high overpopulation and ready for onset of BEC.  The analysis here for the formation of a condensate is on a rather general thermodynamic ground. It is important to further investigate a number of  issues, including: how the condensate forms dynamically, how the condensate formation gets affected by varied initial distribution shape, different amount of overpopulation, as well as the longitudinal expansion and momentum space anisotropy, and what will happen once the inelastic processes are included. These will be addressed later.  

\subsection{From uni-scale to the separation of hard and soft scales}

One more important observation about the initial gluon system is that there is only one scale i.e. the saturation scale $Q_{\rm s}$. It is useful to introduce two scales for characterizing a general distribution: a soft scale $\Lambda_{\rm s}$ below which the occupation reaches $f (p<\Lambda_{\rm s}) \sim 1/\alpha_{\rm s} >>1$ and a hard cutoff scale $\Lambda$ beyond which the occupation is negligible $f (p>\Lambda) << 1$. For initial glasma distribution one has the two scales overlapping $\Lambda_{\rm s} \sim \Lambda \sim Q_{\rm s}$. In contrast for a very weakly coupled thermal gas of gluons one would have the soft scale $\Lambda_{\rm s}^{th} \sim \alpha_{\rm s} T $ and the hard scale $\Lambda^{th} \sim T$ well separated by the coupling $\alpha_{\rm s}$. There is a quite general reason for that. The thermalization is a process of maximizing the entropy (with the given amount of energy). The entropy density for an arbitrary distribution function is given by $s \sim \int_{\vec p} \left[(1+f)\, ln(1+f)-f \, ln (f)\right]$. Obviously the highly occupied modes $f>>1$ contributes only order $\sim \hat{o}(1+lnf)$ to this integration, parametrically the same as those modes with $f\sim 1$. Therefore with the total energy constrained, it is much more beneficial to have a wide window in $\Lambda_{\rm s} <\vec p < \Lambda$ with $f\sim 1$. With this quite general argument, we see that the evolution from glasma to thermal plasma must be accompanied by the separation of scales from the initial uni-scale $\Lambda_{\rm s} / \Lambda \sim 1$ to the final separation $\Lambda_{\rm s} / \Lambda \sim \alpha_{\rm s}$.

\section{Kinetic approach and the scaling solutions}

With the above general insights about the evolution in overpopulated glasma, it is tempting to demonstrate these more explicitly and quantitatively. To do that we have developed a kinetic approach assuming dominance of $2 \leftrightarrow 2$ elastic process \cite{BLM}. The kinetic description is feasible even for the highly occupied regime as discussed in \cite{Mueller:2002gd}.  In the small-angle approximation one can derive the following transport equation:  
\begin{eqnarray} \label{eq_transport}
 {\mathcal D}_t f(\vec p) = \xi \left(\Lambda_{\rm s}^2 \Lambda\right)\, \vec{\bigtriangledown} \cdot  \left[ \vec{\bigtriangledown}f(\vec p) + \frac{\vec{p}}{p}\, \left(\frac{\alpha_{\rm s}}{\Lambda_{\rm s}}\right) f(\vec p)[1+f(\vec p)] \right]
\end{eqnarray}
with $\xi$ an order one constant and the two scales $\Lambda$ and $\Lambda_{\rm s}$ introduced and defined as:
\begin{eqnarray}
\Lambda \left( {{\Lambda_{\rm s}} \over {\alpha_{\rm s}}} \right)^2  \equiv (2\pi^2)\, \int {{d^3p} \over {(2\pi)^3}}\, f\left(\vec p\right)[1+f\left(\vec p\right)]   \quad , \quad
\Lambda {{\Lambda_{\rm s}} \over {\alpha_{\rm s}}}  \equiv  (2\pi^2)  \int {{d^3p} \over {(2\pi)^3}}\, \frac{2\, f\left(\vec p\right)}{p}
\label{eq_def2}
\end{eqnarray}
This equation has been derived in a similar manner to the Landau approach for non-relativistic electromagnetic plasma. The difference from the transport equation for relativistic gluon plasma in e.g. \cite{Mueller:1999pi} lies in that our equation has the full nonlinearity in the $f(1+f)$ terms arising from the Bosonic nature of gluons and becoming extremely important in the highly overpopulated case. This approach also has a marked   difference from previously proposed thermalization scenarios (e.g. the ``bottom-up''\cite{Baier:2000sb,Mueller:2005un}) in that the driving force toward thermalization here is purely elastic scatterings aided by the high overpopulation. A more general equation including finite mass in the particle kinematics is also obtained which has much more complicated structure.

\subsection{Scaling solution in a static box}

Let us first discuss possible scaling solution for the distribution function $f(\vec p)$. As one can see there are two time scales in this problem, the time $t$ itself and the scattering time scale from collision rate $t_{\rm sea} \sim 1/(f^2 \alpha_{\rm s}^2)$. It is precisely because the high overpopulation elevates scattering rate from $\hat{o}(\alpha_{\rm s}^2)$ to $\hat{o}(1)$ that it becomes possible for the two time scales to overlap thus allowing for a scaling solution. We assume the following scaling form and examine its time evolution:
\begin{eqnarray}
f(p<\Lambda)  \sim \frac{\Lambda_{\rm s}}{\alpha_{\rm s}} \,  \frac{1}{p} \quad , \quad f(p>\Lambda) \sim 0
\end{eqnarray}
It is characterized by the two scales $\Lambda$ and $\Lambda_{\rm s}$ that evolve with time. Note that this distribution is consistent with the definition of scales in Eq.(\ref{eq_def2}). With this distribution we immediately see that the coupling constant entirely drops out from the transport equation (\ref{eq_transport}) and the scattering time from the collision integral on the RHS simply  scales as
\begin{eqnarray}
t_{\rm sca} \sim \Lambda / \Lambda_{\rm s}^2 
\end{eqnarray}
Again this is to be compared with a thermal QGP case in which $\Lambda\sim T$ and $\Lambda_{\rm s} \sim \alpha_{\rm s} T$ with a parametrically long scattering time $t_{\rm sca} \sim 1 / (\alpha_{\rm s}^2 T)$. For obtaining a scaling solution, the scattering time shall   scale with the time itself, giving the following condition for   the scales:   
\begin{eqnarray} \label{eq_sca}
t_{\rm sca} \sim \Lambda / \Lambda_{\rm s}^2 \sim t  \,\, .
\end{eqnarray}

Now we further consider the constraint from conservation laws. Let us first deal with the system in a static box, i.e. no expansion. We then must have energy conservation, i.e. 
\begin{eqnarray} \label{eq_en}
\epsilon \sim \frac{\Lambda_{\rm s} \Lambda^3}{\alpha_{\rm s}}  =  {\rm constant}
\end{eqnarray}
The particle number also must be conserved, albeit with a possible component in the condensate: 
\begin{eqnarray} \label{eq_ng}
n =  n_g + n_c  \sim \frac{\Lambda_{\rm s} \Lambda^2}{\alpha_{\rm s}}  + n_c  =  {\rm constant}
\end{eqnarray}
The condensate here plays a vital role with little contribution to energy while unlimited capacity to accommodate excessive gluons thus ensuring the conservation of both. 

Finally with the two conditions (\ref{eq_sca},\ref{eq_en}) we obtain the following scaling solution:
\begin{eqnarray}
\Lambda_{\rm s}  \sim Q_{\rm s} \left( \frac{t_0}{t} \right)^{3/7} \quad , \quad  \Lambda  \sim Q_{\rm s} \left( \frac{t_0}{t} \right)^{-1/7} 
\end{eqnarray}
From this solution, the gluon density $n_g$ decreases as $\sim (t_0/t)^{1/7}$, and therefore the condensate density is growing with time, $n_c \sim (Q_{\rm s}^3/\alpha_{\rm s}) [1-(t_0/t)^{1/7}]$. A parametric thermalization time could be identified by the required separation of scales in a thermal QGP $\Lambda_{\rm s} / \Lambda \sim \alpha_{\rm s}$, giving 
\begin{eqnarray}
t_{\rm th} \sim \frac{1}{Q_{\rm s}} \,  \left( \frac{1}{\alpha_{\rm s}} \right)^{7/4} 
\end{eqnarray} 
At the same time scale the overpopulation parameter $n\epsilon^{-3/4}$ also reduces to order one. 
These results for the static box case here and some part of the analysis in \cite{Kurkela:2011ti} agree. 

\subsection{Scaling solution in the expanding case}

Let us then turn to the longitudinally expanding case that would be more relevant for a realistic heavy ion collision. In this case most of the previous analysis remains, but the conservation laws will be manifest differently. With boost-invariant longitudinal expansion, it is known that the total number density will decrease as $n\sim n_0 t_0/t$. The time-dependence of energy density is also known to be dependent upon the momentum space anisotropy, i.e. for a fixed anisotropy  
\begin{eqnarray}
\epsilon \sim \epsilon_0 (t_0/t)^{1+\delta}
\end{eqnarray}
where the anisotropy parameter is defined by $\delta \equiv 	P_L/\epsilon$ (with $P_L=\int_{\vec p} (p_z^2/p) f(\vec p)$ the longitudinal pressure). We will come back in a moment to a discussion as why the system may maintain a fixed anisotropy. We recognize that the $\delta=1/3$ means isotropy and corresponds to the ideal hydrodynamic limit, while the $\delta=0$ corresponds to the free-streaming limit. Combining this condition with that in Eq.(\ref{eq_sca}) we obtain the following scaling solution in the expanding case:
\begin{eqnarray}
\Lambda_{\rm s}  \sim Q_{\rm s} \left( \frac{t_0}{t} \right)^{(4+\delta)/7} \quad , \quad  \Lambda  \sim Q_{\rm s} \left( \frac{t_0}{t} \right)^{(1+2\delta)/7} 
\end{eqnarray}
With this solution, we see the gluon number density $n_g \sim (Q_{\rm s}^3/\alpha_{\rm s}) (t_0/t)^{(6+5\delta)/7}$, and therefore with any $\delta > 1/5$ the gluon density would drop faster than $\sim t_0/t$ and there will be formation of the condensate, i.e. 
$n_c \sim (Q_{\rm s}^3/\alpha_{\rm s}) (t_0/t) [1-(t_0/t)^{(5\delta-1)/7}]$. Similarly a thermalization time scale can be identified through the separation of scales to be
\begin{eqnarray}
t_{\rm th} \sim \frac{1}{Q_{\rm s}} \,  \left( \frac{1}{\alpha_{\rm s}} \right)^{7/(3-\delta)} 
\end{eqnarray} 

Now we discuss the possibility of maintaining a non-negligible fixed anisotropy during the glasma evolution. One profound effect of rapid longitudinal expansion is that it leads to momentum space anisotropy between the longitudinal and transverse momentum distribution even if one starts with perfectly isotropic distribution. It brings in a new drift term on the LHS of the transport equation $\sim - (p_z/t) \partial_{p_z} f(\vec p)$. If there were no other effect the distribution function will shrink in the $p_z$ direction at a rate $1/t$ (i.e. free-streaming) and a growing anisotropy between $p_z$ and $p_{_T}$ would result. There is of course mechanisms tending to restore the isotropy. At very early time this could be the instabilities (see e.g. ~\cite{Mrowczynski:1993qm,Rebhan:2005re,Dumitru:2006pz}). For the kinetic evolution stage we discuss here, the mechanism is simply the scatterings that try to bring back the balance between longitudinal and transverse pressure and compete against the expansion. The crucial point again is that in the highly overpopulated glasma we consider the scattering rate is parametrically order one, $\Gamma \sim \alpha_{\rm s}^2 f^2\sim \hat{o}(1)$. Furthermore for scaling solutions we care about, the collision term that works to restore the isotropy scales as $C[f] \sim \Lambda_{\rm s}^2 / \Lambda \sim 1/t$. We therefore see a quite matched competition: the expansion tends to increase anisotropy like $\partial_t \delta \sim \delta /t$ while the scattering tends to make it relax back to isotropy like $\partial_t \delta \sim -\delta /t$, and it is conceivable that a balance with a fixed anisotropy could be reached and maintained for a long time. This fixed anisotropy could essentially be anywhere between the two extremes (isotropic/free-streaming), and  depends on the initial anisotropy as well as the pre-factor of scattering term (see \cite{Blaizot:2011xf} for detailed analysis).

\subsection{Kinetic evolution toward the onset of condensation}

Of particular interest is to understand how dynamically the condensation occurs, how that onset is influenced by various factors, and how the condensate and gluons co-evolve further toward thermalization. With the derived equation (\ref{eq_transport}) one can in principle numerically solve it. In the case without initial overpopulation, one indeed can show that the system described by this equation evolves all the way to a thermal Bose-Einstein distribution which is the proper fixed point of the collision term. In the overpopulated case, there is however the complication of the condensate formation. As is well known in atomic BEC literature, one has to separately describe the evolution prior to the onset of condensation (with this equation) and the evolution afterwards (with a coupled set of two equations explicitly for condensate and regular distribution). Efforts have been made in deriving these equations for the coevolution of a condensate and regular distribution \cite{BLM}, but here let us focus on the first pre-BEC stage  and investigate how the system approaches the onset of condensation. This stage is solely described by the Eq.(\ref{eq_transport}) and we have numerically solved it for both the static box and the expanding cases. 

While leaving all the detailed results to be reported in \cite{BLM}, let us describe here the main observation. What we find is that the system with high initial overpopulation will quickly develop a particle flux in momentum space toward the infrared region and pile up particles there. The high occupation number at IR (and thus very fast scattering rate) leads to an almost instantaneous local ``equilibrium'' near the origin $\vec p=0$. This local ``equilibrium'' takes the form:  $f(\vec p\to 0) \to  T^*/(p-\mu^*)$
 with some parameters $T^*,\mu^*$ one may tentatively call the local ``temperature'' and ``chemical potential''. This form is by no means a coincidence --- it is a fixed point of the collision term under the approximation $f(1+f)\approx f^2$ (when $f>>1$). With more and more particles being piled up near the origin, the negative local ``chemical potential'' keeps reducing its absolute value and approaches zero i.e. $(-\mu^*) \to 0^+$. This ultimately marks the onset of the condensation. All these have been explicitly seen  r in the numerical solutions. 

With extensive numerical studies for varied conditions, we have reached the following conclusions for the kinetic evolution of the highly overpopulated system.  For the static box case, the system always reaches onset as long as the initial overpopulation parameter $n\epsilon^{-3/4}$ is greater than the thermal value, $n\epsilon^{-3/4}>0.28$, despite any shape of the initial distribution (e.g. glasma versus Guassian) or any initial anisotropy. 
  The evolution toward onset is still robust in the expanding case: if starting with isotropic distribution, the critical initial overpopulation is shifted mildly to be about $n\epsilon^{-3/4}>0.40$. 
  In the expanding case with initial anisotropy: more longitudinal pressure $P_L>\epsilon/3$ will shift the critical overpopulation to smaller value, while less longitudinal pressure $P_L<\epsilon/3$ will shift it to larger value yet only very slightly even for large anisotropy. 
We therefore see that the link from initial overpopulation toward onset of condensation, in the present kinetic evolution, is a very robust one. We notice that very strong evidences for the formation of Bose condensate have been reported for similar thermalization problem in the classical-statistical lattice simulation of scalar field theory \cite{Epelbaum:2011pc,Gelis:2011xw,Berges:2012us}. The case for non-Abelian gauge theory is more complicated and still under investigation \cite{Berges:2011sb,Berges:2012ev,Berges:2012mc,Kurkela:2012hp,Schlichting:2012es}. A side comment on some of these studies is that usually the coupling $\alpha_{\rm s}$ used in there is several orders of magnitude smaller than one, which may open a wide window between $f\sim 1/\alpha_{\rm s}$ and $f\sim 1$ and in particular a possible ``turbulent'' regime $1<< f << 1/\alpha_{\rm s}$ in between the deep IR and UV. It is a question whether that would survive when $\alpha_{\rm s}$ becomes close to realistic values.

\section{Discussion on the inelastic processes}

This section is devoted to a discussion on the effect of inelastic processes which have 
not been taken into account in the preceding results. Of particular importance is their 
implication for the formation of a condensate. From the viewpoint of equilibrium thermodynamics, 
it is quite obvious that: first, with inelastic processes violating particle number conservation, 
there is no equilibrium chemical potential and there will be no condensate in the equilibrium state; 
second, as argued in e.g. , the inelastic processes (say $2\to 3$ and $3\to 2$) have a much 
faster rate than the $2\to 2$ elastic scattering rate for a weakly coupled quark-gluon plasma in thermal equilibrium \cite{Arnold:2006fz}. 
For the thermalization problem we consider, however,  these points simply do not apply as the 
pre-equilibrium matter is far from equilibrium and needs a different analysis on the roles of the 
inelastic processes. As we will see, in the highly overpopulated glasma even with the presence 
of inelastic processes there could be the dynamical formation of a transient condensate. 

\subsection{The rate of inelastic processes}

As already seen in the previous analysis of the elastic scattering rate, with the high overpopulation $f\sim 1/\alpha_{\rm s}$, the power counting would be rather different from the equilibrium case.  
The inelastic particle production or annihilation processes, e.g. an $n \rightarrow m$ process with $n\neq m$, will contribute new terms to  the collision integral on the right hand side of the transport
equation. For the  $n \rightarrow m$ process, its contribution can be estimated as follows: the vertices contribute a factor $\alpha_{\rm s}^{n+m-2}$ while  the distribution functions  contribute a
factor of $(\Lambda_{\rm s}/\alpha_{\rm s})^{n+m-2}$  (one factor for each distribution, except the one whose momentum one
is following, also noting that the products containing $n+m$ factors $f$
cancel between the gain and loss terms); kinematically the overall infrared singularity \cite{Mueller:2005un} after being cutoff by the Debye mass gives a factor $\sim (1/m_{\rm D}^2)^{n+m-4}\sim (1/\Lambda_{\rm s} \Lambda)^{n+m-4}$, while the remaining momentum integral is dominated by $\Lambda$ scale and leads to a factor $\sim \Lambda^{n+m-5}$ on dimension ground. Quite remarkably one ends up with the inelastic rates $\Gamma_{\rm in} \sim \Lambda_{\rm s}^2/\Lambda$ that have all the couplings cancelled out and that parametrically coincide with the elastic scattering rate $\Gamma_{\rm el} \sim 1/t_{\rm scat} \sim \Lambda_{\rm s}^2/\Lambda$. A few comments follow from this qualitative analysis: first the rate estimates with high overpopulation are quite different from that in equilibrium; second, since the elastic and inelastic rates turn out to be at the same parametric order, a quantitative computation (e.g. numerically solving transport equation with both contributions) would seem necessary for a detailed understanding of their competition and the consequence for the condensate formation; finally, since the time scales for them are scaling in the same way $1/\Gamma_{\rm in} \sim 1/ \Gamma_{\rm el} \sim t$, including the effects of inelastic scattering does not
change the scaling behavior for $\Lambda_{\rm s}$ and $\Lambda$.

\subsection{Possible dynamical balance with the presence of inelastic processes}

While the scaling behavior could remain unchanged with the presence of inelastic processes, the condensate 
evolution however will necessarily be modified. As aforementioned ultimately the inelastic processes 
will eliminate any condensate entirely when reaching equilibrium state, and the question of interest here 
is rather the possible  formation of a transient condensate existing for a considerable time window and 
bearing influence on the dynamics of the system. A quantitative answer is still being pursued by developing 
a detailed numerical analysis, but let us first make a qualitative reasoning here.  As already shown by studying the BEC onset 
in the small  angle approximation to the transport equation for elastic
processes, the condensate forms through a feeding-down source term from a  $1/p$ behavior near $p = 0$ that develops rapidly in the overpopulated case. Now the inelastic terms will contribute a sink term decreasing the condensate. Since the elastic and inelastic processes evolve on the same time scale $\sim t$, it is plausible that a balance may be reached with a nonzero condensate surviving for long. 

To demonstrate this point, let us consider the expanding case without particle number conservation.  We assume the effect of number changing processes may be incorporated by modify the evolution equation for total number density in the following way:
\begin{eqnarray} \label{eq:density_expansion}
 \partial_t n + \frac{n}{t} = \frac{\partial{(t n_c)}}{t dt} + \frac{\partial{(t n_g)}}{t dt} =  - \xi \frac{n_c}{t} \,.
\end{eqnarray}
with $\xi>0$ being a dimensionless parameter from inelastic processes representing their strength. In the above it is essentially assumed that the dominant decrease of particle number comes from processes involving condensate particles. Note that $n=n_c+n_g$ with $n_g$ given from the previous scaling solution. The $n_g$ drops faster than $1/t$ and therefore feeds into the condensate, competing with the decay term on the right hand side. The condensate density, therefore different from the analysis in the number conserving scenario, will now be
\beq \label{eq:condensate_inelastic}
n_c\sim \frac{Q_s^3}{\alpha_s}\, \frac{A}{A-\xi} \,\left( \frac{t_0}{t} \right)^{1+\xi} \left[1 -\left( \frac{t_0}{t} \right)^{A-\xi}    \right].
\eeq
The constant $A$ above is simply defined by $A\equiv (5\delta-1)/7$ for convenience. This solution arises from the dynamical balance of elastic/inelastic processes and shows  interesting features   :\\
1) for $\xi < A$ or equivalently $5\delta-1>7\xi$ we see that at large time one has $n_c\sim \frac{Q_s^3}{\alpha_s}\, \frac{A}{A-\xi} \, ({t_0\over t})^{1+\xi}$ while for gluons $n_g \sim \frac{Q_s^3}{\alpha_s}\,({t_0\over t})^{1+A}$ which implies the condensate dominates the total density;\\
2) for $\xi > A$ or equivalently $5\delta-1<7\xi$ we still have a {\em positive} condensate density and at large time it is simply $n_c \sim \frac{Q_s^3}{\alpha_s}\, \frac{A}{\xi-A} \, ({t_0\over t})^{1+A}$ which drops at the same rate as, and therefore proportional to the density of gluons $n_g \sim \frac{Q_s^3}{\alpha_s}\,({t_0\over t})^{1+A}$;\\
3) in either case, we have found a robust condensate lasting over the evolution even with the presence of the effects from inelastic processes.

\subsection{A subtle point about inelastic processes }

Finally we discuss a subtle point regarding the effects of inelastic processes. We will argue that even in certain cases the individual inelastic rate could be fast, the changing rate for the system's particle number may not be as fast due to the cancellation of the forward and backward going processes. Let's consider for example the $2\to3$ and $3\to2$ processes involving 5 gluons with 2 collinear gluons both in either initial or final states \cite{Mueller:2005un}, i.e. $1,2\to 3,4\parallel 5$ and inverse $3,4\parallel 5 \to 1,2$ with particles $4,5$ begin the collinear ones with splitting fraction $z$ . Note that particle 1 is special in the transport approach as we are ``looking at'' it and not integrating over it. We want to examine the most singular pieces in the IR, i.e. $z\to 0$ (noting $z\to 1$ is a symmetric limit with similar result). In such limit, we have approximately: $f_4(1-z)\to f_4$, $1+f_4(z)\to f_4(z)$ (assuming $f(k\to 0)>>1$), and splitting function $P_gg(z)\to 2C_A/z$. So the collision term in this limiting region factorizes to
\begin{eqnarray}
C_{1,2\leftrightarrow 3,4\parallel 5} \sim C_{1,2\leftrightarrow 3,4} \times \int_{z'\to 0} dz' \frac{2C_A}{z'} f(z') 
\end{eqnarray} 
Note in the above we have made a change of variable $z' = z p_4 $ which is feasible as long as the limit $z\to 0$ is dominant. 
Evidently there is a singularity if $f(p)$ is of power form at small $p$, i.e. $f= \left(\frac{\Lambda}{\omega(p)}\right)^{\gamma}$ for $p\to 0$ and mass (original or medium-given) is zero, as discussed in e.g. \cite{Arnold:2006fz}. Naively one would then conclude 
that the inelastic process in the above factorized limit is infinitely fast and able to eliminate any excessive gluons in no time. 
There is however a subtle point here: the changing rate of particle number is {\em not the same} as the rate of each individual processes. That is to say even though there is a singular piece in each of the above left-going and right-going processes,  there is NO infinitely rapid change of particle number. This can be seen as follows --- the particle number density's changing rate from transport equation is simply 
\begin{eqnarray}
\frac{d n}{dt} = \int_{p_1} C_{1,2\leftrightarrow 3,4\parallel 5} \sim \left[\int_{p_1}  C_{1,2\leftrightarrow 3,4} \right] \times \int_{z'\to 0} dz' \frac{2C_A}{z'} f(z')
\end{eqnarray}
The part in the $[\,\, ]$ must vanish as the $2\leftrightarrow 2$ kernel conserves particle number!   This arises from the symmetric cancelation in the factorized $2\leftrightarrow 2$ part which becomes relevant in the factorable IR limit. One may imagine adding more and more collinear legs to get various inelastic processes --- but in the factorized limit they may presumably  all be ``attachments'' to the factorized basic elastic $2\leftrightarrow 2$ piece which conserves particle number. It could be that summing all these inelastic splitting processes can be effectively absorbed into  the overall coefficient of the elastic process.

\section{Implications on phenomenology }

In various current phenomenological studies in heavy ion collisions there is the ``gap'' between the initial conditions from colliding nuclei 
 and the onset of hydrodynamical evolution \cite{Dusling:2011rz,Schenke:2012wb}, which contributes a significant (if not the largest) part of the theoretical uncertainty for any comparison with experimental data. An improved understanding of the thermalization process will certainly envision new studies on pre-equilibrium phenomenology, e.g. possible contributions to collective flow, jet quenching, as well as electromagnetic emissions in the glasma.  
 
\begin{figure}
	\begin{center}
		\includegraphics[width=6.6cm]{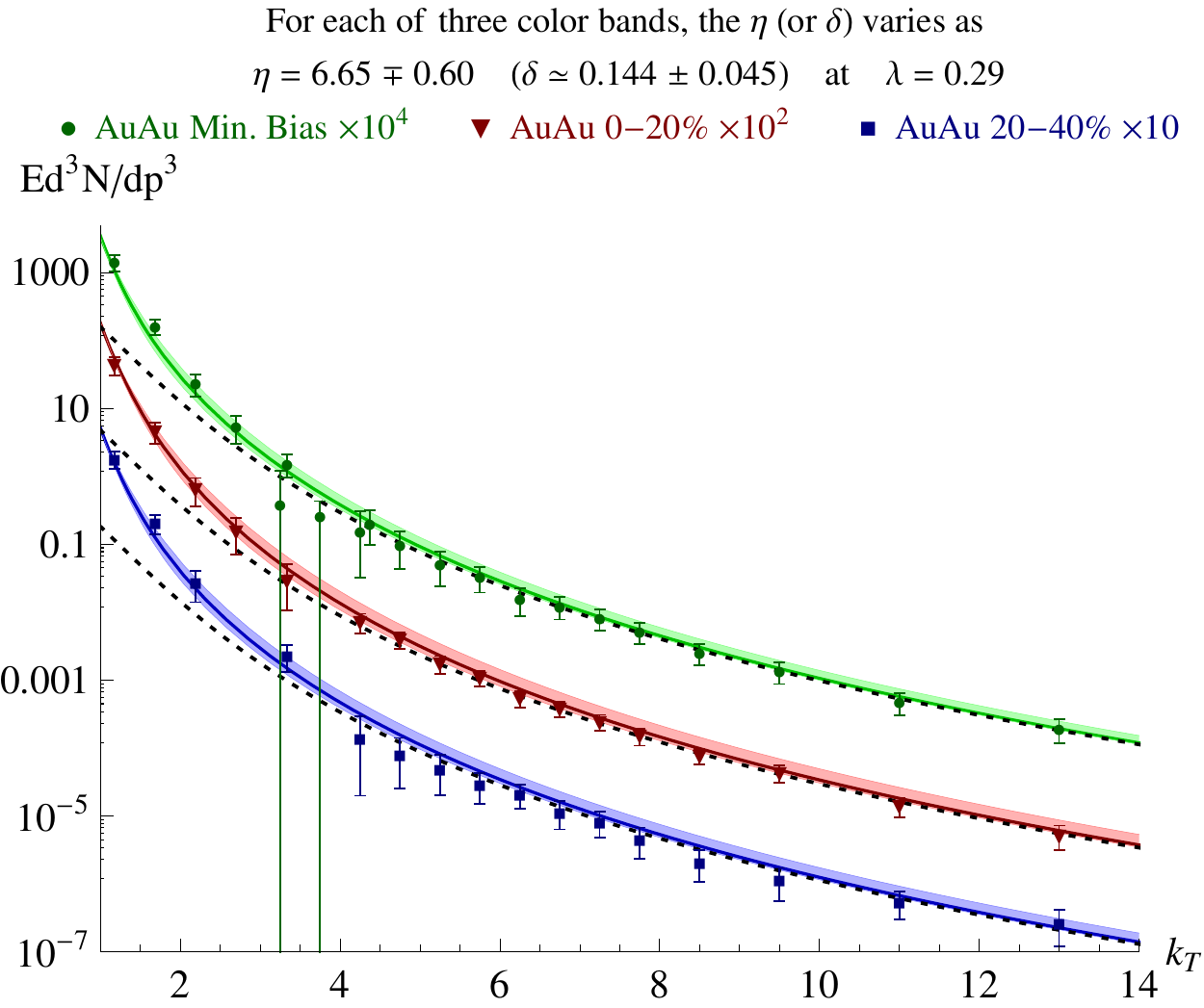} \hspace{0.5cm}
		\includegraphics[width=7.2cm]{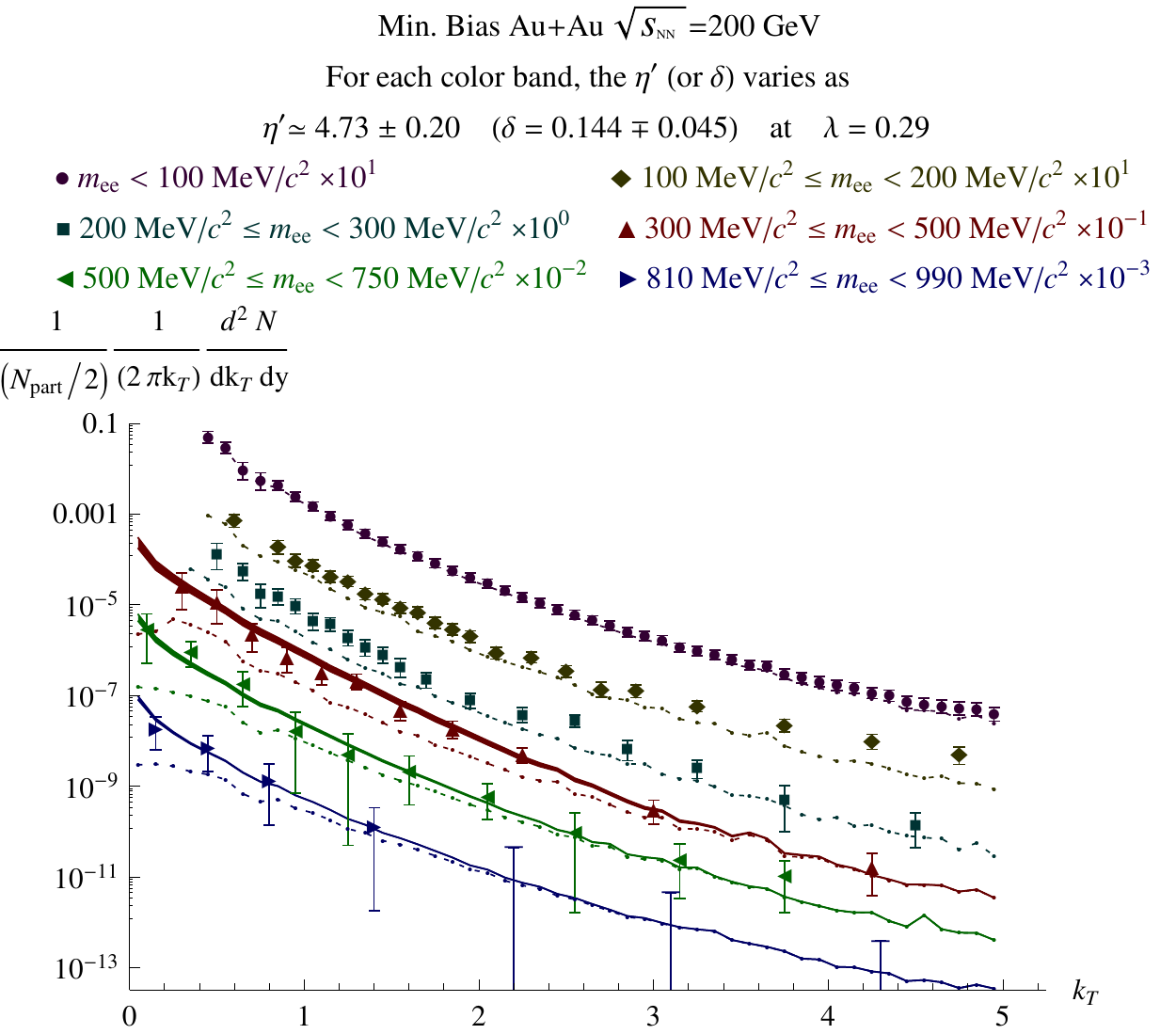}
		\caption{  The $k_T$ spectra of photon (left) and dilepton (right) production including emissions from the pre-equilibrium matter on top of convention sources. The dashed lines are contributions from conventional sources, while the color bands represent the total production including pre-equilibrium emissions, and the data points are from PHINIX measurements (see \cite{Chiu:2012ij} for details). }	\label{EM}
	\end{center}
\end{figure}

Based on the proposed scaling solutions for the glasma evolution \cite{Blaizot:2011xf}, a first attempt has been made in \cite{Chiu:2012ij} to examine the contributions to photon and dilepton production from the glasma. This idea appears interesting particularly due to the well-known ``excess'' in PHENIX measurements which seem not describable by including contributions from various conventional sources. For the photon production that involves (at least) one gluon leg, there is the $f_g\sim 1/\alpha_{\rm s}$ enhancement from the high gluon occupation as compared with the usual thermal case with $f_g\sim 1$. The relevant scale for photon production in glasma is the (evolving) scale $\Lambda$ which is somewhat harder than the scale $T$ of the later thermal QGP. As we can see from Fig.\ref{EM}(left) for photon $k_T$ spectra, indeed in the $k_T\sim$ few $\rm GeV$ region there can be a considerable glasma contribution that makes comparison with data well when added on top of other sources. For the dilepton production in glasma, a possibly important new mechanism may arise from the condensate, i.e. three condensate gluons fusing via a quark loop into virtual photon then dilepton, as suggested and qualitatively analyzed in \cite{Chiu:2012ij}. The key point is that the condensate in glasma has high density $\sim 1/\alpha_{\rm s}$ and low scale $\sim (\Lambda \Lambda_{\rm s})^{1/2}$ which appears to be just right for producing low-mass dileptons in large aggregate. As shown in Fig.\ref{EM}(right), when contribution from this new mechanism is added on top of the background contributions from harmonic cocktail as well as the charmed mesons, the comparison with PHENIX data seems to be very encouraging.  

There are also other interesting pre-equilibrium phenomenology to be fully explored. How much early time flow could be generated during the thermalization process? An answer to this question would be very important for  the goal of extracting transport coefficients of the thermalized QGP. To study this problem in the glasma evolution described here, one needs to incorporate the inhomogeneity of the saturation scale based on realistic nuclear geometry. The dependence of scales $\Lambda$ and $\Lambda_{\rm s}$ on transverse position $\vec r_\perp$ will lead to transport of particles along transverse directions and a full 3D (numerical) kinetic solution may be able to quantify the pre-equilibrium flow.   Could a hard probe lose considerable amount of energy in the evolving glasma? It seems very likely, in particular the high gluon occupation (and the condensate) may well Bose-enhance the soft gluon emissions from a penetrating jet. Such possible pre-equilibrium energy loss is a crucial unknown bearing particular importance for efforts in quantifying the geometric observables of jet quenching as pointed out in e.g. \cite{Jet_Geometry}. Efforts will be needed toward reducing this uncertainty based on the proposed glasma evolution.

\section{Summary}

In summary, we have identified   the most salient feature of the pre-equilibrium gluonic matter (the glasma), namely the high initial overpopulation, and based on that we have proposed a possible scenario with scaling solutions for the glasma evolution  toward a thermalized quark-gluon plasma  in heavy ion collisions.   The strongly interacting nature (and thus fast evolution) naturally  arises as an {\em emergent property} of this pre-equilibrium matter where the  intrinsic coupling is weak but the highly occupied
gluon states coherently amplify the scattering. A possible transient Bose-Einstein Condensate is argued to form dynamically on a rather general ground and the onset of this condensation has been shown numerically in both static and expanding cases based on the kinetic approach we have developed. The role of inelastic processes has been closely examined in the far-from equilibrium situation and the analysis suggests that a transient condensate is still robust. Finally we have discussed pre-equilibrium phenomenology  based on this scenario, in particular the electromagnetic emissions. New avenues are opened for many interesting studies, and we expect further  progress toward quantitative understanding of thermalization. 

\section*{Acknowledgements}

The author thanks his collaborators, J.-P. Blaizot, F. Gelis, L. McLerran, R. Venugopalan, M. Chiu, T. Hemmick, V. Khachatryan, and A. Leonidov. The author is also grateful to the RIKEN BNL Research Center for partial support as well as to the Institute for Nuclear Theory and the organizers of the INT Program ``Gauge Field Dynamics In and Out of Equilibrium''.

\end{document}